\documentclass[12pt,preprint]{aastex}

\shorttitle{Coronal Neutrino Emission}
\shortauthors{Kawabata et al.}

\begin{document}


\title{Coronal Neutrino Emission in Hypercritical Accretion Flows}


\author{R. Kawabata \altaffilmark{1}, S. Mineshige \altaffilmark{1}, and N. Kawanaka \altaffilmark{1}}

\email{kawabata@yukawa.kyoto-u.ac.jp}




\altaffiltext{1}{Yukawa Institute for Theoretical Phisics, Kyoto University, Kyoto 606-8502, Japan}


\begin{abstract}
Hypercritical accretion flows onto stellar mass black holes (BHs) are commonly considered 
as a promising model of central engines of gamma-ray bursts (GRBs).
In this model a certain fraction of gravitational binding energy of accreting matter is deposited 
to the energy of relativistic jets via neutrino annihilation and/or magnetic fields.
However, some recent studies have indicated that the energy deposition rate by neutrino annihilation is somewhat smaller 
than that needed to power a GRB.
To overcome this difficulty, Ramirez-Ruiz \& Socrates (2005) proposed that high energy neutrinos from hot corona above the accretion disk might 
enhance the efficiency of energy deposition.
We elucidate the disk corona model in the context of hypercritical accretion flows.
From the energy balance in the disk and the corona, we can calculate the disk and coronal temperature, $T_\mathrm{d}$ and $T_\mathrm{c}$, 
and neutrino spectra, taking into account the neutrino cooling processes by neutrino-electron scatterings and neutrino pair productions.
The calculated neutrino spectra consist of two peaks; one by the neutrino emission from the disk 
and the other by that from the corona.
We find that the disk corona can enhance the efficiency of energy release but only by a factor of 1.5 or so, 
unless the height of the corona is very small, $H \ll r$.
This is because the neutrino emission is very sensitive to the temperature of emitting region, 
and then the ratio $T_\mathrm{c}/T_\mathrm{d}$ cannot be so large.
\end{abstract}





\section{INTRODUCTION}

Gamma-ray bursts (GRBs) are the most explosive phenomena in the Universe.
Recent development in the observational study of GRBs and their afterglows lead to the understanding of their emission mechanisms
(Piran 2005; M\'{e}sz\'{a}ros 2006 for reviews).
The prompt emissions of GRBs are thought to be emitted from optically thin plasmas with highly relativistic velocity 
in order to avoid the compactness problem (Krolik \& Pier 1991; Lithwick \& Sari 2001).

Some observational studies show jet breaks in their lightcurves, implying the relativistic outflow may be anisotropic, or jet (Rhoads 1999; Sari, Piran, \& Halpern 1999).
GRB prompt emission has an energy $E_{\gamma} \sim 10^{51} \mathrm{ergs} $, if we take into account the beaming of radiations 
(Frail et al. 2001; Bloom, Frail, \& Kulkarni 2003), 
and the kinetic energy of relativistic jet producing GRB is about $E_\mathrm{tot} \sim 10^{52} \mathrm{ergs}$,
if we assume the conversion efficiencies of kinetic energy into radiative energy as about 0.1 (Beloborodov 2000; Guetta, Spada, \& Waxman 2001).

The engines powering such energetic outflows are still unknown.
Some long GRBs show the associations with supernovae type Ic, implying that they originate from the core collapse of massive stars (Bloom et al. 1999; Woosley 1993).
Short GRBs are, on the other hand, recently found to be associated with elliptical galaxies (Gal-Yam et al. 2005),
so they may originate from neutron star (NS) mergers or black hole(BH)-NS mergers (Paczynski 1991; Narayan, Paczynski, \& Piran 1992).
As the result of such energetic phenomena the gravitational energy released in those processes can drive the relativistic jets.
A plausible model of the central engine of GRBs involves an accretion disk 
with extremely high accretion rate ($ \dot{M} \sim 0.01-10 M_{\odot}$) around a stellar mass BH.
Such a flow is called as a neutrino cooled accretion flow or a hypercritical accretion flow (Kohri \& Mineshige 2002).
Recent studies of hypercritical accretion flows have been developed by 
Narayan, Paczynski, \& Piran (1992), Popham, Woosley, \& Fryer (1999), Di Matteo, Perna, \& Narayan (2002, hereafter DPN), Kohri, Narayan, \& Piran (2005) and Kawanaka \& Mineshige (2007).
The radiative cooling is inefficient in such a hot, dense accretion disk with hypercritical accretion rate 
since the optical depth of disk is so high that photons cannot escape thorough the disk within the accretion timescale.
The dominant cooling process of hypercritical accretion flows might be neutrino processes with weak interaction 
and the total energy of neutrino emission reaches some fraction of the rest mass energy of accreting matter (DPN).
The released gravitational energy as neutrino emission can deposit its energy 
to the relativistic fireball via neutrino-antineutrino annihilation in the baryon poor region along the rotational axis.
However some previous work indicated that the energy released via neutrino process might be insufficient for GRB.
The available energy for deposition by neutrino annihilation is only a few tenths of a percent of total energy of emitted neutrinos (Ruffert et al. 1997).
DPN also pointed out that energy deposition rate by neutrino process might be inefficient 
since the accretion flow with a high accretion rate becomes thick for neutrinos and generated neutrinos cannot escape from accreting matter (so called as neutrino trapping). 

Ramirez-Ruiz \& Socrates (2005, hereafter RS) indicated that 
non-thermal neutrinos generated in the hot corona (with temperature $T_\mathrm{c}$) above the hypercritical accretion disk (with temperature $T_\mathrm{d}$) might enhance the energy deposition by neutrino annihilation 
since the deposition rate is proportional to the mean energy of neutrinos and $T_\mathrm{c}$ can be greater than $T_\mathrm{d}$.
They assumed the coronal depth for neutrinos to estimate the coronal temperature and 
concluded that energy deposition rate may increase by some factor 
if a sufficiently thin corona forms.
However they did not calculate the neutrino spectra and mean energy of emitted neutrinos.
Moreover the coronal region they assumed may be too small, 
and thus, the coronal temperature might have been overestimated.

In this paper we elucidate the disk corona model in the context of hypercritical accretion scenario 
and show that the mean energy of emergent neutrinos from hypercritical accretion flows is enhanced 
by the existence of neutrino-thin corona.
We take a reasonable value of the coronal thickness and calculate the emergent spectra of neutrinos from the corona.
In \S 2 our model of disk corona is described with some adopted assumptions. 
We show the numerical method to calculate the neutrino spectra, 
and energy equations to determine the temperature of the corona and the disk are written in \S3.
We consider the absorbed neutrinos by the dense disk, which re-emits thermal neutrinos, to determine the coronal temperature self-consistently, which RS did not take into account.
Then we present our results in \S 4 and discuss the possibility of the corona to enhance the energy deposition rate in \S 5.

\section{OUR MODEL}


We first outline our model in this section.
Originally, disk corona model was developed to explain the non-thermal spectral component of X-ray binaries and active galactic nuclei (AGNs) 
in analogy with the model for the solar corona (see e.g. Shibata, Tajima, \& Matsumoto 1990).
From the fitting to the observed spectra it has become clear that 
$f$, the fraction  of gravitational energy released in diffuse corona, should be almost unity, 
i.e, almost all the liberated gravitational energy should be dissipated in the corona, 
since otherwise the large flux of the non-thermal spectral component, which is compatible to that of the thermal component, cannot be explained (Haardt \& Maraschi 1991).
The magnetic pressure can be amplified up to the equipartition value with the gas pressure by MHD processes, 
such as magnetorotational instability in the accretion disk (Balbus \& Hawley 1998; Machida \& Matsumoto 2003).
Magnetic loops are formed and lifted up by the Parker instability, creating a corona filled with magnetic fields (Galeev, Rosner, \& Vaiana 1979; Stella, L., \& Rosner, R. 1984).
This corona may be heated by magnetic reconnection.
In this way some fraction $f$ of total gravitational energy can be released in the corona.
Liu, Mineshige, \& Shibata (2002) have shown that $f \sim 1$ from the consideration of energy balance in the disk corona system.

Hot corona might also form in hypercritical accretion flows since the accretion disk is highly turbulent (RS).
Thus, following the ordinary disk corona model mentioned above, 
we construct the disk corona model in the context of hypercritical accretion flows for GRBs.
The high energy neutrinos from the hot corona can enhance the energy deposition rate by neutrino annihilation, 
since the annihilation cross section depends on the energy of neutrinos.
The energy deposition rate by neutrino annihilation at per unit volume is (Ruffert et al. 1997)
\begin{eqnarray}
q^+ _{\nu \bar{\nu} \to e^- e^+}= \frac{2 ({C_\mathrm{V}}^2 + {C_\mathrm{A}}^2)}{3 \pi} {G_\mathrm{F}}^2 \  \int d \Omega _{\nu} \int d \Omega _{\bar{\nu}}
      \left( \langle E_{\nu} \rangle + \langle E_{\bar{\nu}} \rangle \right) \ I_{\nu} I_{\bar{\nu}} (1 - \cos \theta)^2 ,
\end{eqnarray}
where
$C_\mathrm{V} = 1/2 + 2 {\sin}^2 \theta _\mathrm{W}$ and $C_\mathrm{A} = 1/2$ for electron type pairs, 
$C_\mathrm{V} = -1/2 + 2 {\sin}^2 \theta _\mathrm{W}$ and $C_\mathrm{A} = -1/2$ for heavy-lepton neutrino pairs, 
$G_\mathrm{F}$ is the Fermi coupling constant,
$\langle E_{\nu} \rangle = \int E I_{\nu}(E) dE/\int I_{\nu}(E) dE$ (or $\langle E_{\bar{\nu}} \rangle = \int E I_{\bar{\nu}}(E) dE/\int I_{\bar{\nu}}(E) dE$) 
is the mean energy of neutrinos (antineutrinos), 
$I_{\nu}$ (or $I_{\bar{\nu}}$) is intensity of emitted neutrinos (antineutrinos), 
and $\theta$ is the collision angle, and $\Omega _{\nu}$ (or $\Omega _{\bar{\nu}}$) is the solid angle of neutrino (antineutrino) emission, 
which depend on the geometry of emitting region.
We take the Weinberg angle, $\theta _\mathrm{W}$, as ${\sin}^2 \theta _\mathrm{W} = 0.23$.
For simplicity, we assume that neutrinos and antineutrinos are all electron type and have the same intensity.
Thus the energy deposition rate is roughly proportional to the mean energy of neutrinos.
If the hot corona emitting neutrinos form, the neutrino spectra, which are generally assumed as thermal, 
may be deformed and have higher mean energy than original value.

The energy released in the accretion disk per unit surface can be calculated by using the standard disk theory 
and is written as (Kato, Fukue, \& Mineshige 1998)
\begin{eqnarray}
Q^+ = \frac{3 G M \dot{M}}{8 \pi r^3 } \left( 1- \sqrt{\frac{r_\mathrm{in}}{r}} \right),
\end{eqnarray}
where $M$, $\dot{M}$, and $r_\mathrm{in}$ is the mass of black hole, the accretion rate, 
and the inner boundary radius of the accretion disks where the torque vanishes, respectively.
We take $M = 3 M_{\odot}$ and $r_\mathrm{in} = 3 r_\mathrm{s} \simeq 3 \times 10^6 \mathrm{cm}$ where $r_\mathrm{s}$ is the Schwarzschild radius.
This dissipated energy is, if the neutrino cooling is efficient, released as neutrino emission. 
We consider the plane-parallel and homogeneous corona with vertical thickness $H$ above the disk (Fig.\ref{fig:corona}).
With conservative views the parameter value $H$ may be comparable to the disk thickness, 
since the most unstable wavelength of Parker instability is nearly scale height in the disk (e.g. Matsumoto et al. 1988) 
and thus, the typical scale of the magnetic field in the corona would be comparable to the scale height.

Neutrino emission from the inner part of the disk dominates over that from the outer part 
and mainly contributes to the heating of relativistic fireball above the disk.
Then, we evaluate the enhancement of mean energy of neutrinos at $r = 4 r_\mathrm{s}$, where the energy dissipation rate reaches nearly a maximum value, 
and we assume that this enhancement is proportional to that of the total energy deposition rate by neutrino annihilation.
We assume that the corona consists of pure relativistic electron-positron plasma 
since the corona may form above the surface of the disk where the density of baryon is much less than that inside of the disk (discussed later).
We also assume that the electrons and positrons are completely thermalized, 
since the timescale of electromagnetic interactions is much shorter than that of weak interactions by many order of magnitude.

As the coronal cooling processes we take into account neutrino reactions as
\begin{eqnarray}
\nu + e^{\pm} \to \nu +e^{\pm} \\
e^- + e^+ \to \nu + \bar{\nu}.
\end{eqnarray}\\
The thermal neutrinos emitted from the disk are up-scattered by hot electrons and positrons in the corona.
The scattered neutrinos have higher energy since the temperature of the corona ($T_\mathrm{c}$) is higher than that of the disk ($T_\mathrm{d}$) 
and, hence, emerged neutrino spectrum is deformed.
The corona is cooled via scatterings of neutrinos by high energy electrons and positrons in the corona.
Some part of scattered neutrinos are re-absorbed and heat the disk, which re-emits thermal neutrinos.
Neutrino pairs produced by annihilation of high energy electrons and positrons in the corona also cool the corona and have high energy spectra.
We neglected the scattering of neutrinos produced by pair process in the corona 
since the energy of neutrinos emitted in the corona is almost the same energy as that of hot electrons 
and since the optical depth of electron-neutrino scattering is much less than unity in the corona.
We also neglected the Fermi blocking effect by background of electrons, positrons and neutrinos.

On the basis of these assumptions, we can calculate $T_\mathrm{d}$ and $T_\mathrm{c}$ by solving energy equations including neutrino processes, 
which are needed to calculate the neutrino spectra (see \S 3.3).




\section{CALCULATIONS OF NEUTRINO SPECTRA}

Hereafter we take the units as $c$=1, $\hbar=1$, and $k$=1.
Calculating the neutrino spectrum emitted from the disk corona system is important when we evaluate the energy deposition rate 
by neutrino annihilation, since the deposition rate is proportional to the neutrino energy.
In this section we present the method to calculate the coronal neutrino spectra and their mean energy.


\subsection{Neutrino-electron Scattering}

We calculated the change of energy of neutrinos up-scattered by hot coronal electrons and positrons.
The differential cross section of neutrino-electron scattering is ('t Hooft 1971; Kneller, McLaughlin, \& Surman 2006)
\begin{eqnarray}
\frac{d \sigma _\mathrm{e\nu}}{d K_\mathrm{e}} &=& \frac{{G_\mathrm{F}}^2 {m_\mathrm{e}}^2}{2 \pi m_\mathrm{e}} \left \{ (C_\mathrm{V}+C_\mathrm{A})^2+(C_\mathrm{V}-C_\mathrm{A})^2(1-\frac{K_\mathrm{e}}{\overline{E}_ \nu})^2 \right. \nonumber \\
                               & & \ \ \ \ \ \ \ \ \ \ \ \ \ \ \ \ \ \ \ \ \ \ \ \ \ \ \ \ \ \ \ \ \ \ \ \ \ \ \ \ \ \ \ \ \left. -({C_\mathrm{V}}^2-{C_\mathrm{A}}^2)\frac{m_\mathrm{e} K_\mathrm{e}}{{\overline{E}_ \nu}^2} \right \},
\end{eqnarray}
where 
$C_\mathrm{V} = 1/2 + 2 {\sin}^2 \theta _\mathrm{W}$ and $\ C_\mathrm{A} = 1/2$ for electron-electron neutrino scattering, 
$C_\mathrm{V} = 1/2 + 2 {\sin}^2 \theta _\mathrm{W}$ and $\ C_\mathrm{A} = -1/2$ for positron-electron neutrino scattering, 
$K_\mathrm{e}$ is the kinetic energy of recoiled electron, and $\overline{E}_ \nu$ is the energy of an incident neutrino in the rest frame of electron.
We do not deal here with the composition of neutrino flavors, since the detailed composition depends on the structure of the disk.
We assume that the emitted neutrinos from the disk are all electron type since nucleon pair capture is a dominant neutrino process in the disk.

The mean free path, $\lambda$, of neutrinos in the corona for neutrino-electron scattering is then (Landau \& Lifshits 1976)
\begin{eqnarray}
\lambda ^{-1} = \frac{2}{(2 \pi)^3} \int \sigma _\mathrm{e\nu} (1- \mu \beta) f_\mathrm{e} d^3 p, 
\end{eqnarray}
where $\beta$, $p$, $f_\mathrm{e}$ are the velocity, momentum, and the distribution function of electrons, respectively,
and $\mu$ is cosine of the angle between the direction of neutrino and electron.
We assume that the corona consists of pure electron positron plasma in thermal equilibrium by electromagnetic process 
so that the distribution function of pair should be Fermi-Dirac type, $f_\mathrm{e} = [\exp({\sqrt{p^2 + m_\mathrm{e}^2}/T}) + 1] ^{-1}$.
In the relativistic limit, where $T_\mathrm{c} \gg m_\mathrm{e}$, the total coronal depth for neutrino-electron scattering is
\begin{eqnarray}
\tau &\equiv& \frac{H}{\lambda} \nonumber \\
     &\simeq& 0.27 \left( \frac{H}{10^6 \mathrm{cm}} \right) \left( \frac{E_\nu}{10 \mathrm{MeV}} \right) \left( \frac{T_\mathrm{c}}{10 \mathrm{MeV}} \right)^4,
\end{eqnarray}
where $E_\nu$ is the energy of incident neutrino in the laboratory frame.
Hence, higher energy neutrinos have more chances to collide with coronal electrons than lower ones.

We calculate the neutrino spectra scattered by coronal electrons and positrons 
by using Monte Carlo method, following  Pozdnyakov, Sobol, \& Sunyaev (1977) and Liu, Mineshige, \& Ohsuga (2003).
We first set the weight of neutrino $w_0=1$ for a given thermal neutrino with energy, $E_\nu$, which has the Fermi-Dirac distribution.
We calculate the probability for neutrinos passing through the corona, $P_0 = \exp (-\tau/\cos \alpha)$, 
where $\alpha$ is the angle between neutrino direction and the $z$-axis which is perpendicular to the coronal plain.
Then $w_0 P_0$ is the transmitted portion of neutrinos 
and the remaining $w_1=w_0(1-P_0)$ is the portion of neutrinos scattered at least once.
Let $w_{n} = w_{n-1} (1 - P_{n-1})$ be the portion of neutrinos experiencing the $n$-th scattering.
We continue the calculation until the weight, $w_n$, becomes sufficiently small.
Repeating the same procedures for sufficiently large number of neutrinos, 
we can calculate emergent spectra by collecting neutrinos going upward through the coronal surface at $z=H$,
while downward neutrinos crossing the lower boundary are re-absorbed by the disk body, thereby heating the disk.
%

\subsection{Neutrino-antineutrino Pair Production}

We also calculate the neutrino pair emission by weak interaction in the pair plasma.
The cross section for pair process by weak interaction is (Dicus 1972; Yakovlev et al. 2001)
\begin{eqnarray}
\sigma _{e^- e^+ \to \nu \bar{\nu}} v = \frac{{G_\mathrm{F}}^2}{12 \pi} \frac{{m_\mathrm{e}}^4}{E_1 E_2} \left[ ({C_\mathrm{V}}^2 + {C_\mathrm{A}}^2) \left(1 + 3 \frac{P_1 \cdot P_2}{{m_\mathrm{e}}^2} +2 \frac{(P_1 \cdot P_2)^2}{{m_\mathrm{e}}^4} \right) \right. \ \ \ \ \ \ \ \ \ \ \ \ \ \ \ & \nonumber \\
                                                            \left. + 3({C_\mathrm{V}}^2 - {C_\mathrm{A}}^2) \left( 1 + 2 \frac{P_1 \cdot P_2}{{m_\mathrm{e}}^2} \right) \right], &
\end{eqnarray}
where $C_\mathrm{V} = 1/2 + 2 {\sin}^2 \theta _\mathrm{W}$ and $C_\mathrm{A} = 1/2$ for pair production of electron types, 
$C_\mathrm{V} = -1/2 + 2 {\sin}^2 \theta _\mathrm{W}$ and $C_\mathrm{A} = -1/2$ for pair of $\mu$ and $\tau$ neutrinos, 
$E_1$(or $E_2$) and $P_1$(or $P_2$) are the energy and four momentum of electron (positron), respectively, and $v$ is the relative velocity of the pair.
The neutrino pair emissivity is given as
\begin{eqnarray}
q_{e^- e^+ \to \nu \bar{\nu}} = \frac{4}{(2 \pi)^6} \int d^3 p_1 d^3 p_2 (E_1 + E_2) \sigma _{e^- e^+ \to \nu \bar{\nu}} v f_1 f_2 .                            \label{eq:pair-emissivity-pre}
\end{eqnarray}
We assume that the distribution functions of electrons and positrons are Fermi-Dirac type with the same temperature, $T_c$.
In the relativistic limit where $T_\mathrm{c} \gg m_\mathrm{e}$, the total emissivity becomes
\begin{eqnarray}
q_{e^- e^+ \to \nu \bar{\nu}} = 1.39 \times 10^{34} \left( \frac{T_\mathrm{c}}{10 \mathrm{MeV}} \right)^9 \mathrm{\ ergs \ cm^{-3} \ s^{-1}}                   \label{eq:pair-emissivity}
\end{eqnarray}
and the mean energy of neutrinos is 
\begin{eqnarray}
\langle E \rangle &=& \frac{\int d^3 p_1 d^3 p_2 {E_1}^2 \sigma _{\nu\nu} v f_1 f_2}{\int d^3 p_1 d^3 p_2 E_1 \sigma _{\nu\nu} v f_1 f_2} \nonumber \\
                  &\simeq& 5.1 \ T_\mathrm{c} .
\end{eqnarray}
A half of the emitted neutrinos pass through the corona upwardly and the remaining half are absorbed and heat the disk since emission of pair process is isotropic.
The emissivity, equation (\ref{eq:pair-emissivity}) includes all types of neutrinos but we assume that the emitted neutrinos by pair-process are electron type only 
when considering heating of the disk by downward neutrinos and the energy deposition to the relativistic fireball by upward neutrinos.
This is a reasonable assumption in doing simple estimation, 
since the ratios of the emissivity of each type of neutrinos is, from equation (\ref{eq:pair-emissivity-pre}), 
0.70:0.15:0.15 for $\nu_\mathrm{e},\nu_{\mu},\nu_{\tau}$.

We also calculated the neutrino spectra from pair process by Monte Carlo method.
The reaction rate of electron positron annihilation by weak interaction is given as
\begin{eqnarray}
dR_{e^- e^+ \to \nu \bar{\nu}} = \frac{4}{(2 \pi)^6} \sigma _{e^- e^+ \to \nu \bar{\nu}} v \ f_1 d^3 p_1 \ f_2 d^3 p_2 .                       \label{reaction}
\end{eqnarray}
Thus, for given energy of electrons and positrons the reaction weight, $w$, is proportional to $\sigma _{e^- e^+ \to \nu \bar{\nu}} v$.
The differential cross section of neutrino pair process in the frame of the center of mass is (Misiaszek, Odrzywotek, \& Kutschera 2006),
\begin{eqnarray}
\frac{d \sigma _{e^- e^+ \to \nu \bar{\nu}}}{d \Omega} &\propto& | \mathcal{M} |^2 \nonumber \\
                          &=& 8 {G_\mathrm{F}}^2 \left[ ({C_\mathrm{V}} - {C_\mathrm{A}})^2 P_1 \cdot Q_1 P_2 \cdot Q_2 \right. \nonumber \\
                          & & \ \ \left. + ({C_\mathrm{V}} + {C_\mathrm{A}})^2 P_2 \cdot Q_1 P_1 \cdot Q_2 + {m_\mathrm{e}}^2 ({C_\mathrm{V}}^2 - {C_\mathrm{A}}^2) Q_1 \cdot Q_2 \right], 
\end{eqnarray}
where $P_1$(or $P_2$) and $Q_1$(or $Q_2$) are the four momentum of electron (positron) and neutrino (antineutrino), respectively, 
and $| \mathcal{M} |^2$ is the amplitude of pair process.
We determine the emitted energy of neutrino pair with reaction weight $w$ from the differential cross section formula 
and calculate the neutrino spectra from pair process.


\subsection{Energy Balances in the Corona and the Disk}

We calculate the temperature of the corona and that of the disk consistently by solving their energy balances simultaneously and by calculating the neutrino spectra (see Fig.\ref{fig:energy}). 
We assume that the fraction $f$ of the total energy is dissipated in the corona and remaining part, $1-f$, in the disk, neglecting the advection of energy.
The energy balance in the disk is 
\begin{eqnarray}
(1-f) Q^+ + Q_{\mathrm{ref}} + \frac{1}{2} q_{e^- e^+ \to \nu \bar{\nu}} H= \frac{7}{8} \sigma {T_\mathrm{d}}^4,                                                     \label{eq:disk-balance}
\end{eqnarray}
where we assumed that the corona is thin for neutrinos emitted by pair process so that we can neglect the absorptions by the inverse pair process.
On the other hand, the energy balance in the corona is 
\begin{eqnarray}
f Q^+ + \frac{7}{8} \sigma {T_\mathrm{d}}^4 = Q_{\mathrm{esc}} + Q_{\mathrm{ref}} + q_{e^- e^+ \to \nu \bar{\nu}} H,                                        \label{eq:corona-balance}
\end{eqnarray}
where $Q_{\mathrm{esc}}$ is the cooling rate by the thermal neutrinos from the disk and upward neutrinos scattered by electrons and positrons, 
and $Q_{\mathrm{ref}}$ is that by the downwardly scattered neutrinos, 
which can be evaluated by calculating the scattered neutrino spectra.

If all of the energy is dissipated within the disk, i.e., when $f$=0,
the effective temperature of the disk at $r = 4 r_\mathrm{s}$ is 
\begin{eqnarray}
T_\mathrm{d,0} = 4.2 \left( \frac{M}{3 M_{\odot}} \right)^{- 1/2} \left( \frac{\dot{M}}{1 M_{\odot} \mathrm{s}^{-1}} \right)^{1/4} \ \mathrm{MeV},
\end{eqnarray}
and the mean energy of thermal neutrinos from the disk is $ \langle E \rangle _0 \simeq 4.1T_\mathrm{d,0}$.
With given parameters, $f$ and $H$, we can solve these equations (\ref{eq:disk-balance}) and (\ref{eq:corona-balance}) iteratively for $T_\mathrm{d}$ and $T_\mathrm{c}$, 
together with the calculated neutrino spectra.
Then, we evaluate the enhancement of the mean energy of neutrinos from the corona, comparing with that of thermal one, $4.1T_\mathrm{d,0}$.

\section{RESULTS}

\subsection{Neutrino-electron Scattering}

First we consider the effect of scattering alone as a cooling mechanism of the corona 
in analogy with the coronae of standard accretion disks in X-ray binaries and AGNs.
From equations (\ref{eq:disk-balance}) and (\ref{eq:corona-balance}), and neglecting the pair process, we can calculate the temperature of the corona and the disk by 
calculating the change of neutrino energy $Q_{\mathrm{esc}}$ and $Q_{\mathrm{ref}}$ by Monte Carlo method iteratively.
Figure \ref{fig:scattering} shows the coronal and disk temperatures for $\dot{M} = 1 M_{\odot} \mathrm{s}^{-1}$ as functions of $f$.
A thinner corona (with smaller $\tau$) has a higher temperature for a given $T_\mathrm{d}$, since the cooling rate of the corona by scatterings is roughly written as 
\begin{eqnarray}
f Q^+ \sim \frac{7}{8} \sigma {T_\mathrm{d}}^4 \times \tau \times \frac{T_\mathrm{c}}{T_\mathrm{d}},
\end{eqnarray}
where we assume that the scattered neutrino has nearly the same energy of that of electrons,
and that $\tau$ is proportional to the thickness of the corona.

We then show the neutrino spectra where $f$ is close to unity (Fig.\ref{fig:scat-spectra}, $left$).
The neutrino spectra have two components: thermal neutrinos from the disk and scattered neutrinos in the hot corona.
Nearly a half of the scattered neutrinos go downward and are absorbed by the disk, which generate soft thermal neutrinos, 
and the remaining half pass upward producing a high energy spectrum.
Once thermal neutrinos are scattered, they acquire energy nearly equal to that of the coronal electrons, 
since the energy of neutrino is too high ($T_\mathrm{d} \gg m_\mathrm{e} $) to be scattered elastically.
The high energy part of emergent neutrino spectra mainly contribute to neutrino heating by neutrino annihilation.
If $H$ is on the same order of $r$, in contrast, the the coronal temperature is not significantly high compared with disk temperature.
Thus, the effect of up-scatterings is small and the spectrum does not show two humps so clearly.

We also calculate the mean energy of neutrinos from neutrino spectra 
and the amplification of neutrino mean energy is shown in Figure \ref{fig:scat-spectra} ($right$).
With conservative value of coronal thickness, $H/r \sim 0.1-1$, the amplification factor of mean energy is about a factor of two.

\subsection{Neutrino Pair Production}

In the next case we take into account the effect of cooling by neutrino pair production. 
Neglecting the scattering, the temperature of the corona and the disk can be solved analytically.
The temperature of the corona and the disk at $r = 4 r_\mathrm{s}$ are
\begin{eqnarray}
T_\mathrm{c} &=& 5.6 \ f^{1/9} \left( \frac{M}{3 M_{\odot}} \right)^{-1/3} \left( \frac{\dot{M}}{1 M_{\odot} \mathrm{s}^{-1}} \right)^{1/9} \left( \frac{H}{r} \right)^{-1/9} \ \mathrm{MeV} \label{eq:pair-Tc} \\
T_\mathrm{d} &=& 4.2 \ \left(1-\frac{f}{2} \right)^{1/4} \left( \frac{M}{3 M_{\odot}} \right)^{-1/2} \left( \frac{\dot{M}}{1 M_{\odot} \mathrm{s}^{-1}} \right)^{1/4} \ \mathrm{MeV} \label{eq:pair-Td},
\end{eqnarray}
{
and the mean energy of emergent neutrinos is $ \langle E \rangle \simeq 4.1T_\mathrm{d} (1 - f/2) + 5.1 T_\mathrm{c} f/2$.

Including the effect of scatterings we also solved equations (\ref{eq:disk-balance}) and (\ref{eq:corona-balance}), and evaluated $T_\mathrm{d}$ and $T_\mathrm{c}$.
Figure \ref{fig:scat+pair} shows the temperatures in the cases with $\dot{M} = 1 M_{\odot} \mathrm{s}^{-1}$.
The effect of neutrino electron scattering is negligible and the behavior of temperature is well described by equations (\ref{eq:pair-Tc}) and (\ref{eq:pair-Td}).

The emerged neutrino spectra with neutrino electron scatterings and pair process are shown in Figure \ref{fig:spectra} ($left$).
Note that the effect of scatterings is so small that scattered neutrinos make a small hump at the electron energy, 
which is overlaid by the spectrum of pair process.
In the case where $f \sim 1$ half of the energy dissipated in the corona is reprocessed in the disk as thermal neutrinos with lower temperature 
and the half emitted by pair process with higher temperature pass through the neutrino-thin corona.
Figure \ref{fig:spectra} ($right$) shows the amplification of neutrino mean energy, including the scattering and pair process.
If we take $H/r = 0.1-1$, the mean energy of neutrinos is enhanced by a factor of about 1.5.
A thinner corona becomes hotter and emits neutrinos with higher energy because the neutrino emission occurs in a small dissipation region as the cooling process.

\subsection{$\dot{M}$ Dependence}

There is a hope that $T_\mathrm{c}$ may increase with an increase of $\dot{M}$ so that the energy amplification factor could increase as $\dot{M}$ increases.
Therefore, we also calculated the amplification of the mean energy of neutrinos in the cases where 
$\dot{M} = 0.1 M_{\odot} \mathrm{s}^{-1}$ and $10 M_{\odot} \mathrm{s}^{-1}$.
From Figure \ref{fig:mdot-depend} we see that the mean energy of neutrinos from the corona with high accretion rate tends to be slightly enhanced 
since the $\dot{M}$ dependence of the coronal temperature is weak.
In the case with $\dot{M} = 0.1 M_{\odot} \mathrm{s}^{-1}$ the mean energy of neutrinos is enhanced by a factor two with $H = 0.1 r$ 
but the neutrino luminosity (and also the absolute value of mean energy) is small.
On the other hand when $\dot{M} = 10 M_{\odot} \mathrm{s}^{-1}$, the mean energy is enhanced only by 1.2 
even though we neglect the effect of neutrino trapping.
If we consider the neutrino trapping, the enhancement will be smaller.
Thus, there is practically no improvement in the neutrino energy, even if $\dot{M}$ is larger.

\section{SUMMARY AND DISCUSSION}

\subsection{Brief Summary}
Magnetically heated corona above the accretion disk may emit non-thermal, high energy neutrinos,
which, in principle, would lead to an enhancement of the energy deposition rate by neutrino annihilation.
The improvement is, however, only by a factor of two or so when $H/r \gtrsim 10^{-3}$.
Hence, the energy deposition might still be insufficient to energize GRBs unless $H/r$ is extremely small.
We see that higher energy of neutrinos is expected if the corona form in a thiner region compared to the disk scale height.

We also see that the effect of the scattering between electrons and neutrinos is negligible compared with the cooling by pair process.
The ratio of the energy loss rate of the corona by electron-neutrino scatterings to the emission rate from the corona by pair process is approximately
\begin{eqnarray}
\frac{(7 \sigma {T_\mathrm{d}}^4 /8) \times (T_\mathrm{c} / T_\mathrm{d}) \tau}{q_{\nu \bar{\nu}} H} \simeq 0.1 \left( \frac{T_\mathrm{d}}{T_\mathrm{c}} \right)^4.
\end{eqnarray}
Then, the effect of scattering is negligible at the higher coronal temperature, which is the situation which we are considering.
We can easily understand this ratio by considering the temperature dependences of the two neutrino processes.
The number density of coronal electrons is roughly $n_\mathrm{e} \propto {T_\mathrm{c}}^3$ and the cross section for pair process is $\sigma _{e^- e^+ \to \nu \bar{\nu}} \propto {T_\mathrm{c}}^2$, 
thus the emissivity of neutrino pair production is proportional to $T_\mathrm{c} \sigma _{e^- e^+ \to \nu \bar{\nu}} {n_\mathrm{e}}^2 \propto {T_\mathrm{c}}^9$.
On the other hand, the number density of thermal neutrinos is $n_{\nu} \propto {T_\mathrm{d}}^3$ and the cross section for electron neutrino scattering is $\sigma _\mathrm{e\nu} \propto T_\mathrm{d} T_\mathrm{c}$.
Hence, the emissivity of neutrinos scattered by coronal electrons is proportional to $T_\mathrm{c} \sigma _\mathrm{e\nu} n_\mathrm{e} n_{\nu} \propto {T_\mathrm{d}}^4 {T_\mathrm{c}}^5$, 
where we assume that the energy of scattered neutrino is roughly $T_\mathrm{c}$.
From these simple estimations, we understand that the emission of pair process dominates that of electron neutrino scatterings by a factor of $(T_c/T_d)^4$.

From Figure \ref{fig:scat+pair}, we see that the temperature of the disk does not decrease so much even if $f$ increases.
This is can be understood in the following way: 
since about a half of energy of emitted neutrinos in the corona is absorbed and heats the disk, 
more than a half of the gravitational energy, $Q^+$, is emitted as thermal neutrinos from the disk.
Hence the disk temperature decreases by only a factor of $0.5^{1/4} \sim 0.84$ 
even if $f \sim 1$ (the case in which most of the gravitational energy is dissipated in the corona), 
compared with the the cases without disk coronae.
%
%
%
%
%

We also investigated the dependence of the mean energy of neutrinos on the accretion rates.
The disk temperature $T_\mathrm{d,0}$ is proportional to $\dot{M}^{1/4}$ since the dissipative energy is proportional to $\dot{M}$, 
while the coronal temperature is proportional to $\dot{M}^{1/9}$ if the pair process dominates.
Hence the ratio $T_\mathrm{c}/T_\mathrm{d,0}$ is proportional to $\dot{M}^{-5/36}$.
Though the neutrino luminosity is proportional to $\dot{M}$, even if we neglect the advective energy by neutrino trapping in the disk, 
the enhancement of mean energy of neutrinos is small at high accretion rates.
Thus we cannot expect significant enhancement of neutrino heating by high energy neutrinos even at high accretion rates.
If we take into account of neutrino trapping, we need replace the fraction of energy dissipated in the disk, $1-f$, by $1-f-f_\mathrm{adv}$ 
where $f_\mathrm{adv}$ is the fraction of advective energy.
We could neglect the advective energy when $f \sim 1$ since the energy is transported via magnetic process to the surface of the disk 
on the dynamical timescale, which is shorter than the neutrino diffusion time (Ohsuga et al 2002).
This will lead to improvement of the result of DPN even though the enhancement of the mean energy of neutrinos may still be small.

\subsection{Confinement of Coronal Plasma}
The corona might be formed by the energy release of magnetic fields in a thin region above the disk.
What can then confine the magnetic corona?
If we consider the hydrostatic balance in the corona, the thickness of the corona becomes 
\begin{eqnarray}
H \sim \frac{c_\mathrm{s}}{\Omega_\mathrm{K}} \sim \sqrt{\frac{2}{3}} \left( \frac{r}{r_\mathrm{s}} \right) r \gtrsim r \ \ \ \ (r>r_\mathrm{s}),
\end{eqnarray}
where we assumed the sound speed in the corona as $c_\mathrm{s} = 1/\sqrt{3}$.
Hence the corona cannot be gravitationally bound but expands, 
unless the radial position of the corona is very close to the Schwarzschild radius.
The ratio of the cooling time, $t_\mathrm{cool} \sim 11 a {T_\mathrm{c}}^4 /4 q_{e^- e^+ \to \nu \bar{\nu}} $, 
to the free expansion time of the corona, $t_\mathrm{exp} \sim H/c_s$, is 
\begin{eqnarray}
\frac{t_\mathrm{cool}}{t_\mathrm{exp}} \sim \frac{22}{7 \sqrt{3}} {\tau _{\nu \bar{\nu}}}^{-1} > 1,
\end{eqnarray}
where $\tau _{\nu \bar{\nu}} = q_{e^- e^+ \to \nu \bar{\nu}} H / 4(7/8) \sigma {T_\mathrm{c}}^4 $ is the neutrino optical depth of the corona for pair process (RS).
Thus, the corona should expand before the coronal plasma is cooled by neutrino emission since we now consider the neutrino-thin hot corona, 
and some processes are needed to confine the coronal plasma in a thin region.

The magnetic pressure could be amplified up to the equipartition value of the gas pressure in the disk 
and the amplified magnetic fields will be lifted up in the corona (RS).
Such strong magnetic fields in the corona might be able to confine the coronal plasma 
if the magnetic pressure is higher than that of relativistic particles in the corona.
Since the gas pressure in the disk is roughly $P_\mathrm{gas} \sim 10^{30} \mathrm{ergs \ cm^{-3}}$ for $\dot{M} = 1 M_{\odot} \mathrm{s}^{-1}$ (DPN), 
the magnetic fields, whose pressure is comparable to $P_\mathrm{gas}$, could confine the corona if $T_\mathrm{c} \lesssim 10 \mathrm{MeV}$, i.e. $H \gtrsim 10^{-3} r$ in our model (see Fig. \ref{fig:scat+pair}).
Our estimates above include some uncertainties, e.g. the detailed structure of magnetic fields, 
and we need further studies to see whether such magnetic fields really can confine the coronal plasma.

\subsection{Baryon Contamination in the Corona}
Finally, we discuss on the validity of the assumption that the corona consists of pure pair plasma with neglegible baryons.
In the existence of baryons, neutrino emission by the pair capture process may also cool the corona.
The cooling rate by the pair capture process dominates over that by the pair process if the corona contains sufficient amount of baryons, 
$\rho _{10} > 0.6 T_\mathrm{c,10}^3$, where $\rho _{10}$ is the density of baryons in the unit of $10^{10} \mathrm{g \ cm^{-3}}$ and 
$T_\mathrm{c,10} = T_{c}/\mathrm{10 MeV}$ (Popham, Woosley, \& Fryer 1999).
Thus, for the validity of the assumptions in our model, the baryon density in the corona should be smaller than that in the dense disk, 
whose baryon density is typically $10^{10-11} \mathrm{g \ cm^{-3}}$ for neutrino thick disk (e.g. Kawanaka \& Mineshige 2007).
Since the coronal temperature is $T_\mathrm{c,10} \lesssim 1$ with a moderate thickness of the corona in our model, $H \gtrsim 10^{-3} r$, 
the baryon density in the corona should be less than that in the disk by one or two orders of magnitude.
This condition could be marginally justified.

Moreover the distribution function of electrons in the corona could be deformed from the Fermi-Dirac type of zero chemical potential.
In such a case the neutrino emission by the pair process could be suppressed (Kohri \& Mineshige 2002).
Since the chemical potential of electrons is determined by the amount of baryons in the corona, 
this calculation is beyond the scope of this work and should be done in future works.





\acknowledgments

The authors are grateful to S. Nagataki and Y. Masada for discussions and comments.
The numerical calculations were carried out at YITP in Kyoto University.
This work is supported by a Grant-in-Aid for the 21st Century COE 'Centre for Diversity and Universality in Physics'.

\clearpage



\begin{figure}
\epsscale{.50}
\plotone{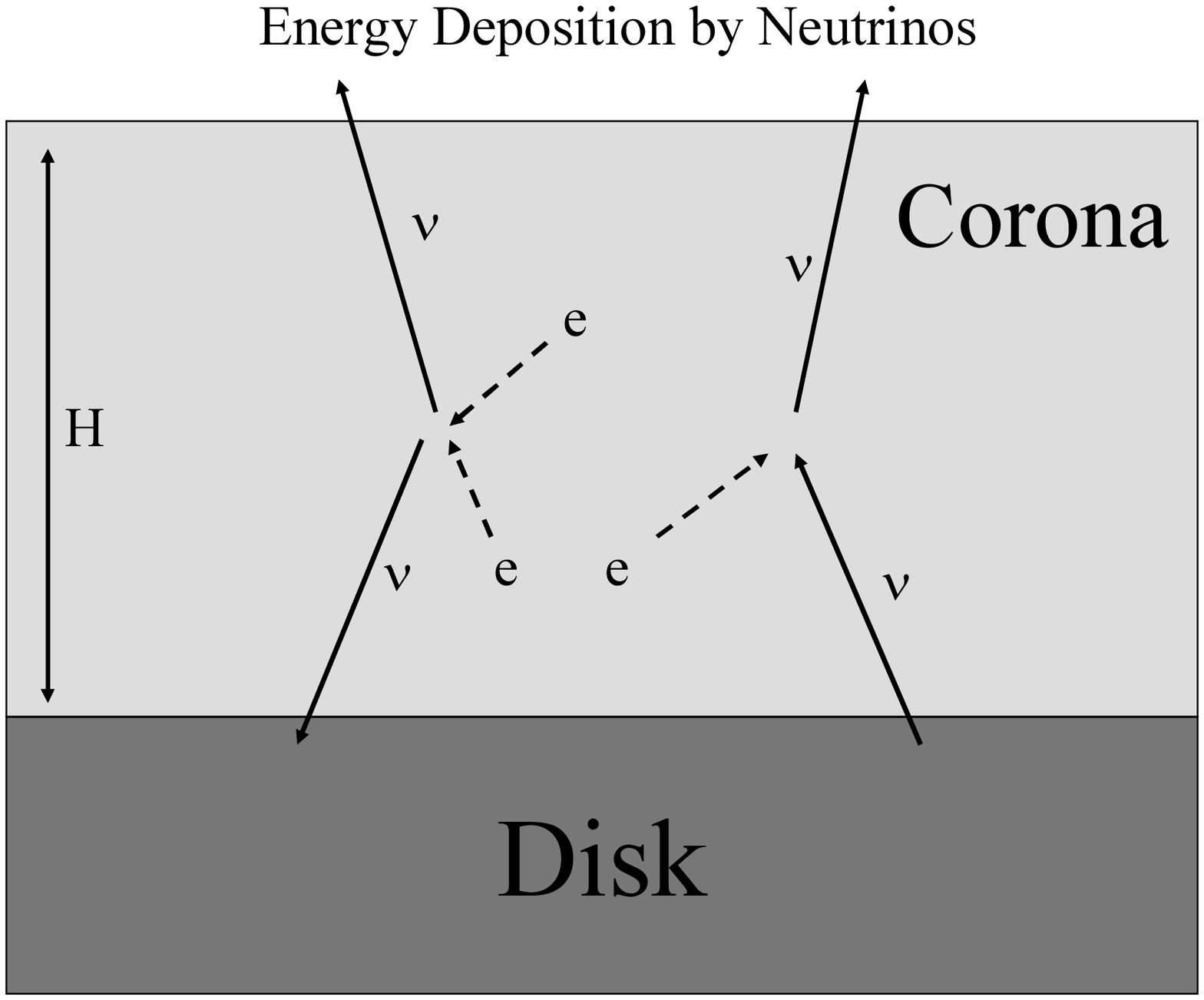}
\caption{The corona model in hypercritical accretion flows.
\label{fig:corona}
}
\end{figure}

\clearpage

\begin{figure}
\epsscale{.50}
\plotone{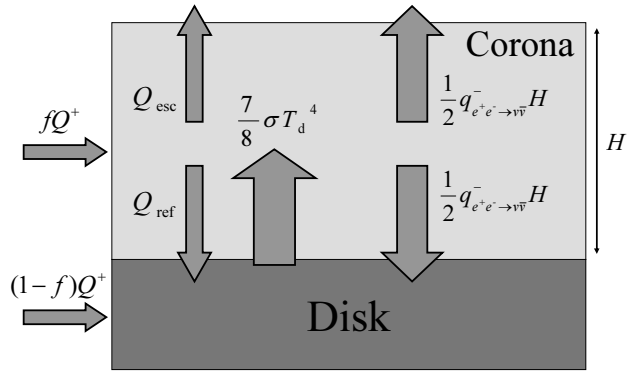}
\caption{Energy balance in the corona and the disk by neutrino processes.
A fraction of liberated energy, $f Q^+$, is dissipated in the corona, and $(1-f) Q^+$ in the disk.
\label{fig:energy}
}
\end{figure}

\clearpage

\begin{figure}
\epsscale{.80}
\plotone{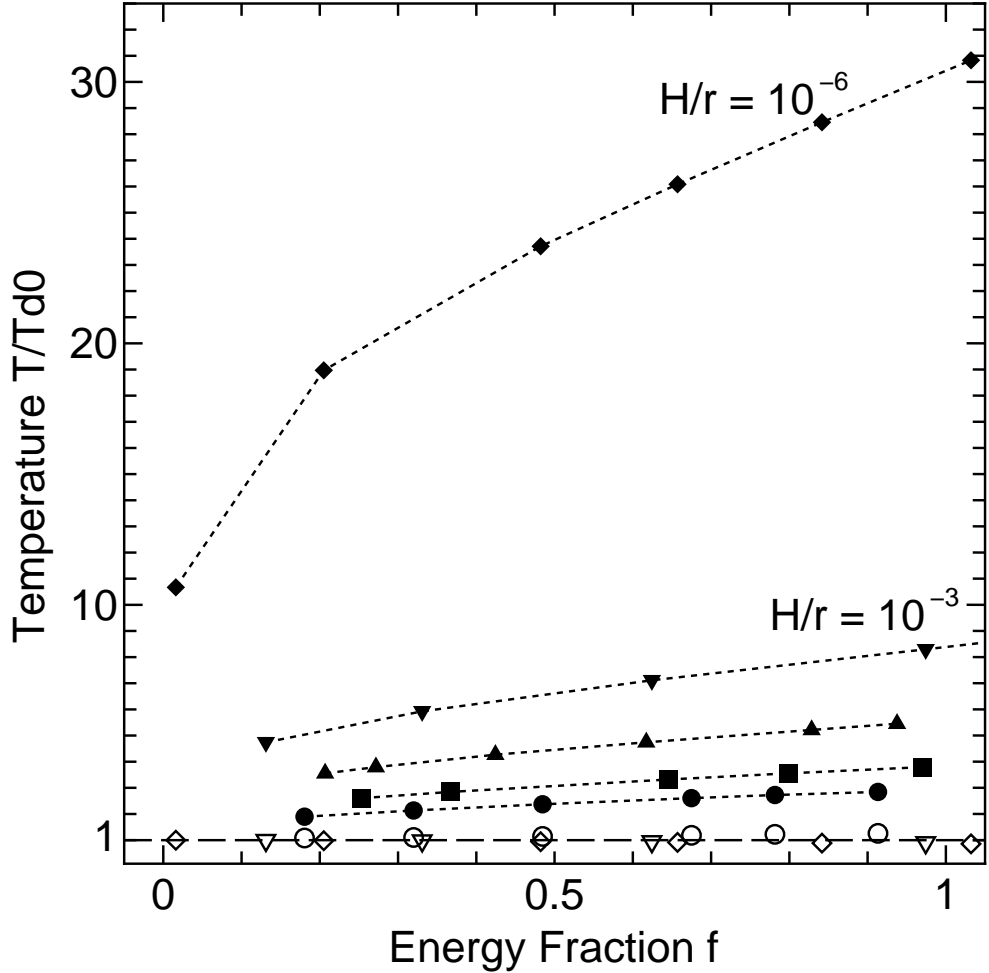}
\caption{Temperatures of the corona, $T_\mathrm{c}$($filled$), and the disk, $T_\mathrm{d}$($open$), as functions of $f$. 
The parameters are $\dot{M} = 1 M_{\odot} s^{-1}$ and the various coronal thicknesses, 
$H/r$=1 ($circle$), $10^{-1}$ ($square$), $10^{-2}$ ($upward \ triangle$), $10^{-3}$ ($downward \ triangle$) and $10^{-6}$ ($diamond$).
Note that $T_\mathrm{c}$ and $T_\mathrm{d}$ are normalized by $T_\mathrm{d,0} = 4.2$ MeV.
\label{fig:scattering}
}
\end{figure}

\clearpage

\begin{figure}
\epsscale{1.0}
\plottwo{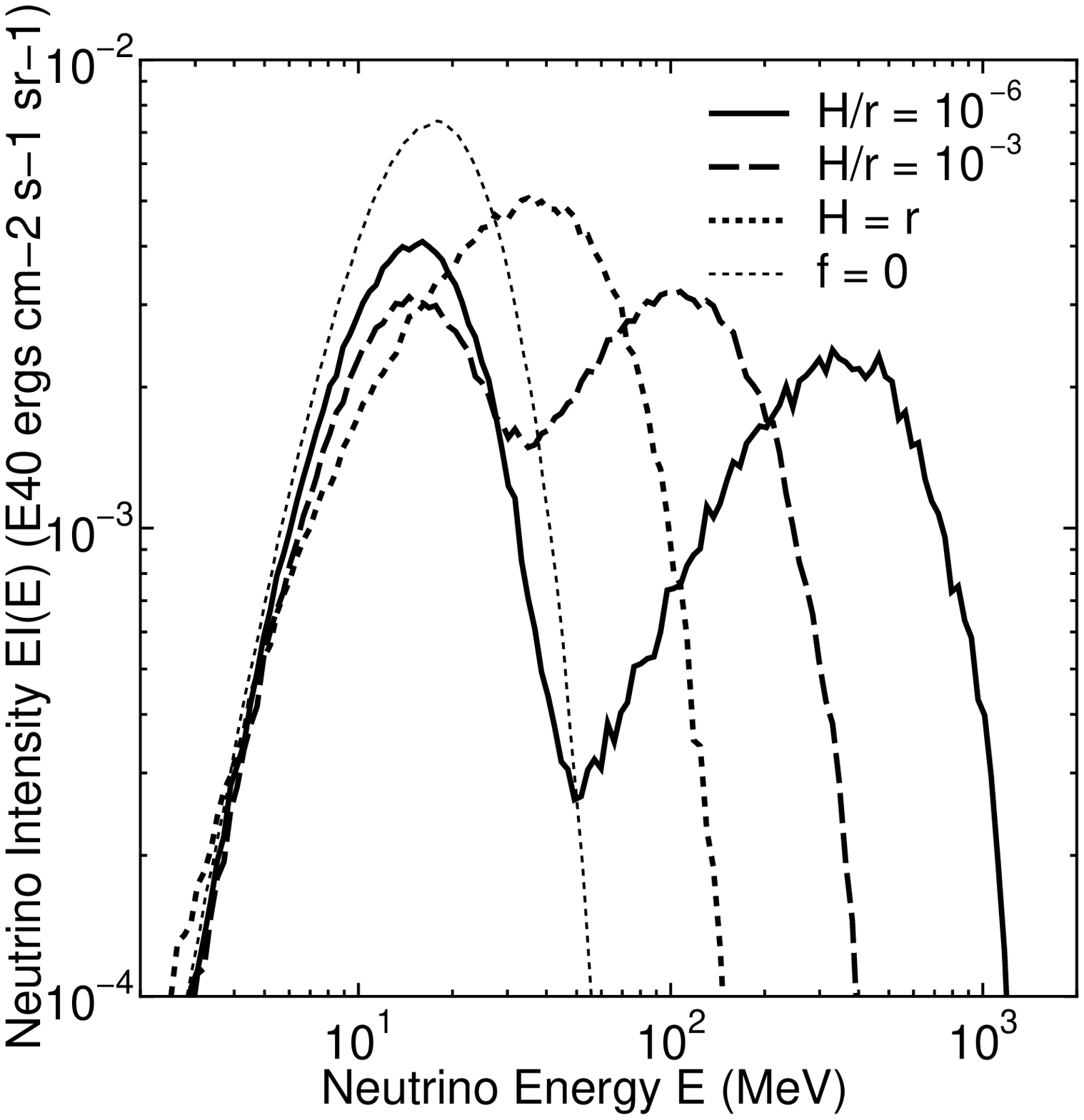}{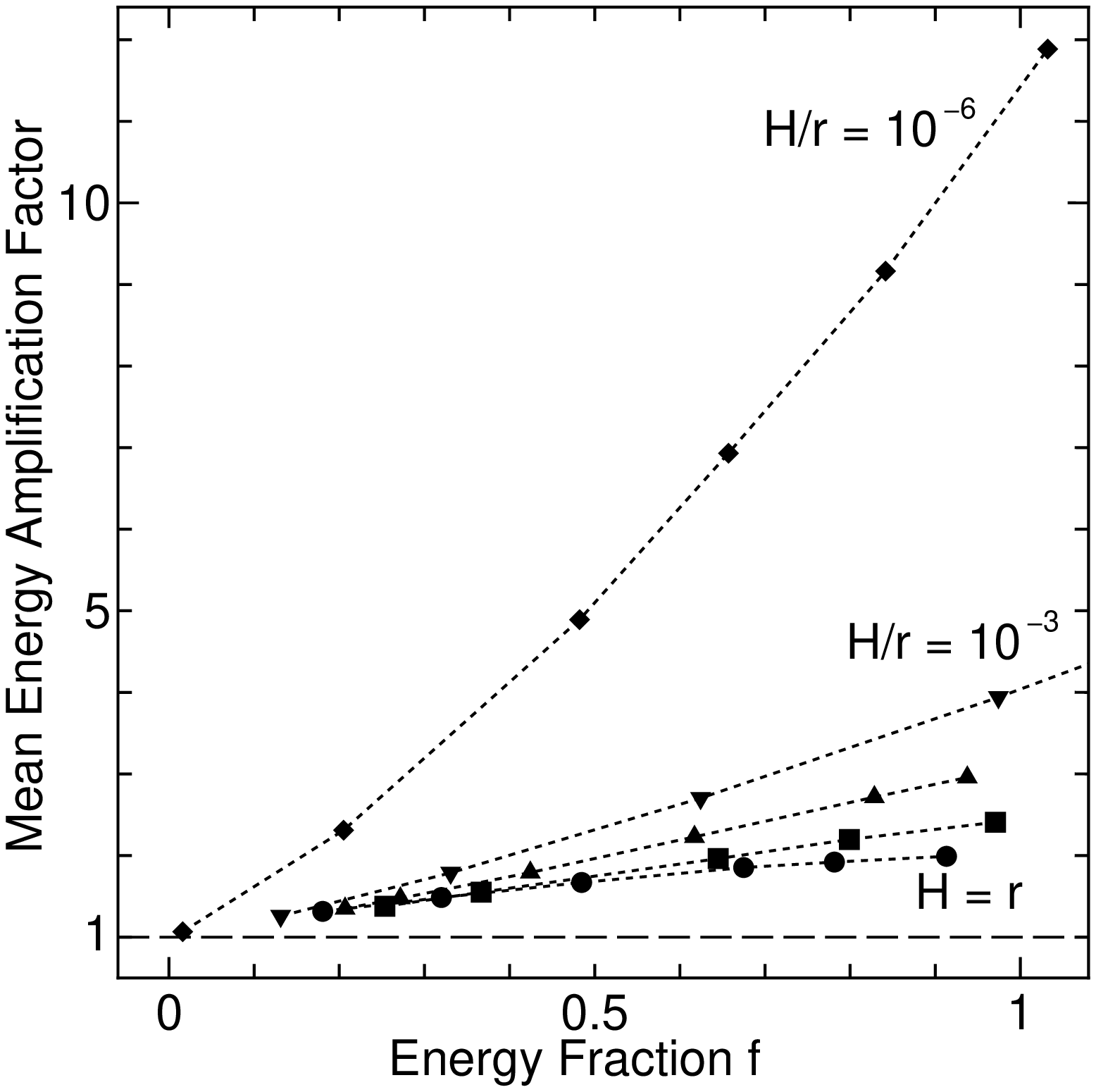}
\caption{$Left$: Neutrino spectra emerged from the corona at $r = 4 r_\mathrm{s}$. 
Spectra of thermal neutrinos with $T_\mathrm{d,0}$ ($dot \ line$) and coronal neutrino spectra 
with $(H/r, f)$=(1, 0.91) ($short \ dashed \ line$), $(H/r, f)$=($10^{-3}$, 0.97) ($dashed \ line$) and $(H/r, f)$=($10^{-6}$, 0.84) ($solid \ line$), respectively.
$Right$: Same as Fig.\ref{fig:scattering} but for the mean energy of emerged neutrinos normalized by $ \langle E \rangle _0 \simeq 17$ MeV, 
the value of the case with $f=0$ (no corona).
\label{fig:scat-spectra}
}
\end{figure}

\clearpage

\begin{figure}
\epsscale{.50}
\plotone{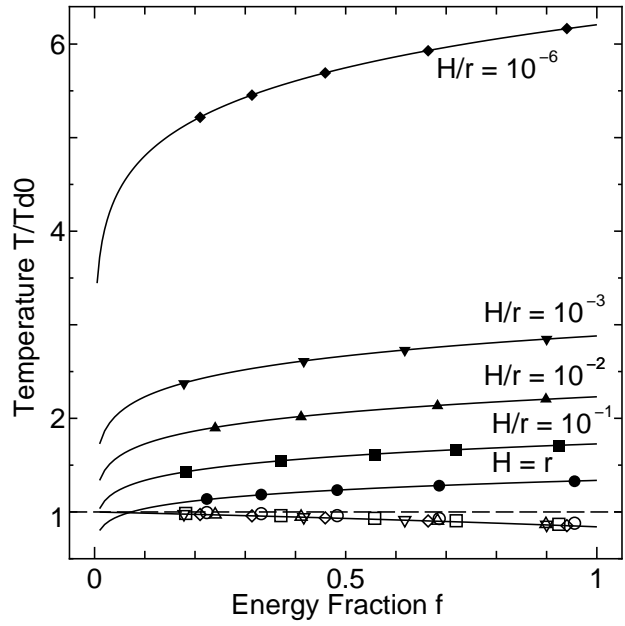}
\caption{Same as Fig.\ref{fig:scattering} but for the cases where the neutrino pair process is also considered.
The solid line shows the temperatures where the neutrino-electron scatterings are neglected.
\label{fig:scat+pair}
}
\end{figure}

\clearpage

\begin{figure}
\epsscale{1.0}
\plottwo{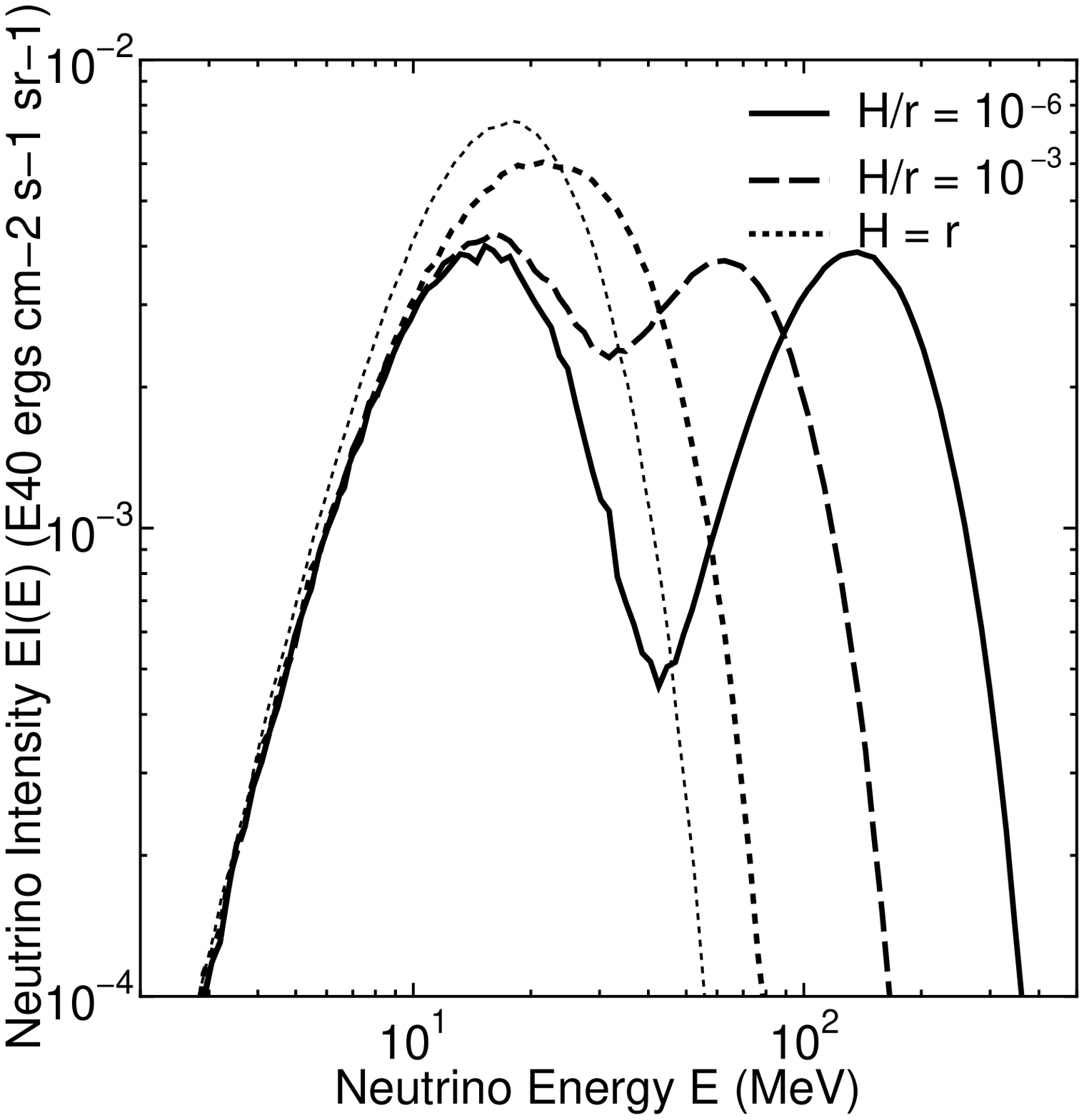}{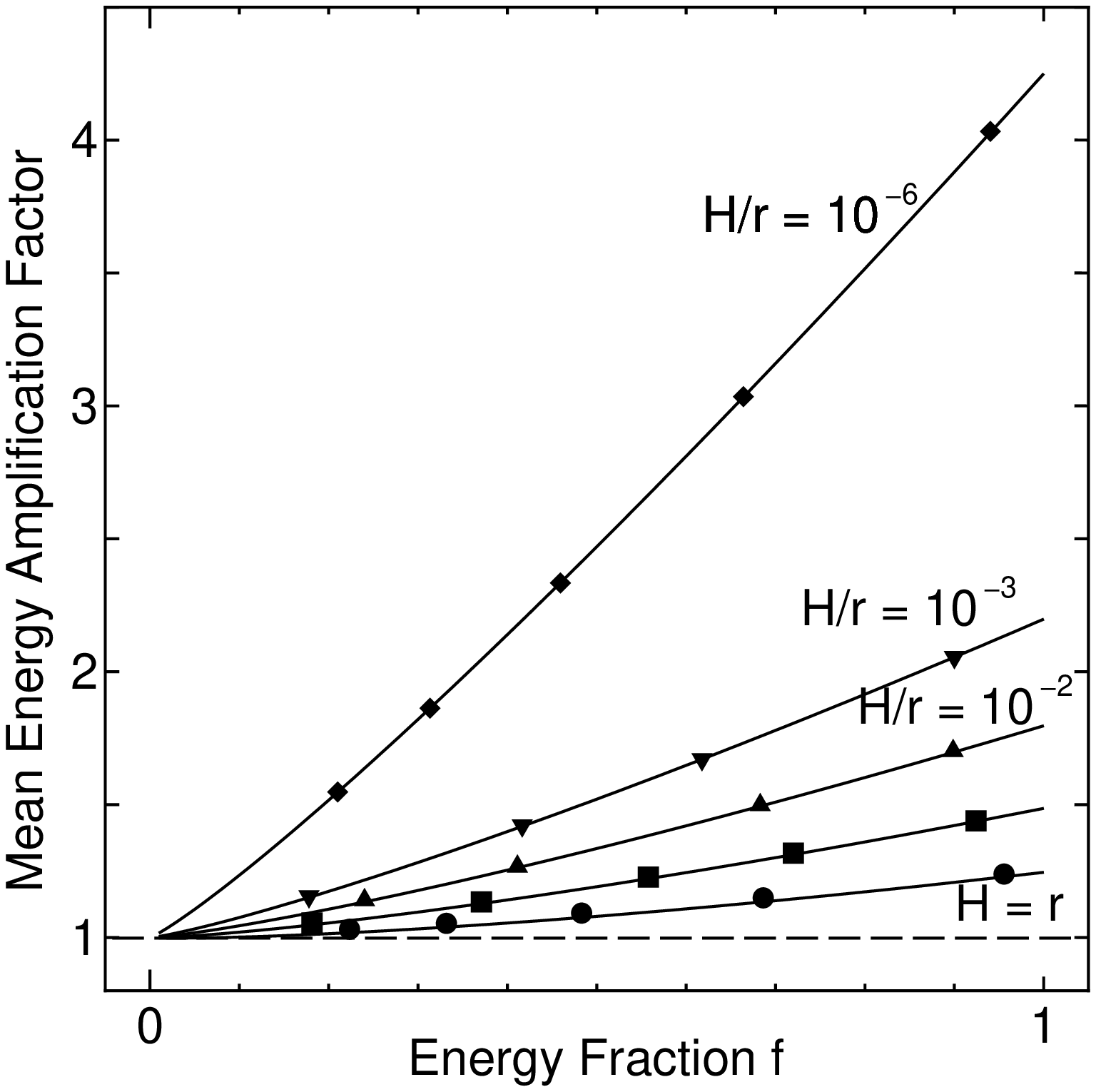}
\caption{$Left$: Same as Fig.\ref{fig:scat-spectra} but for the case with neutrino pair production.
The parameters are $(H/r, f)$=(1, 0.96) ($short \ dashed \ line$), $(H/r, f)$=($10^{-3}$, 0.90) ($dashed \ line$) and $(H/r, f)$=($10^{-6}$, 0.94) ($solid \ line$).
$Right$: The amplification factor of the neutrino mean energy, same as Fig.\ref{fig:scattering}
\label{fig:spectra}
}
\end{figure}

\clearpage

\begin{figure}
\epsscale{1.0}
\plottwo{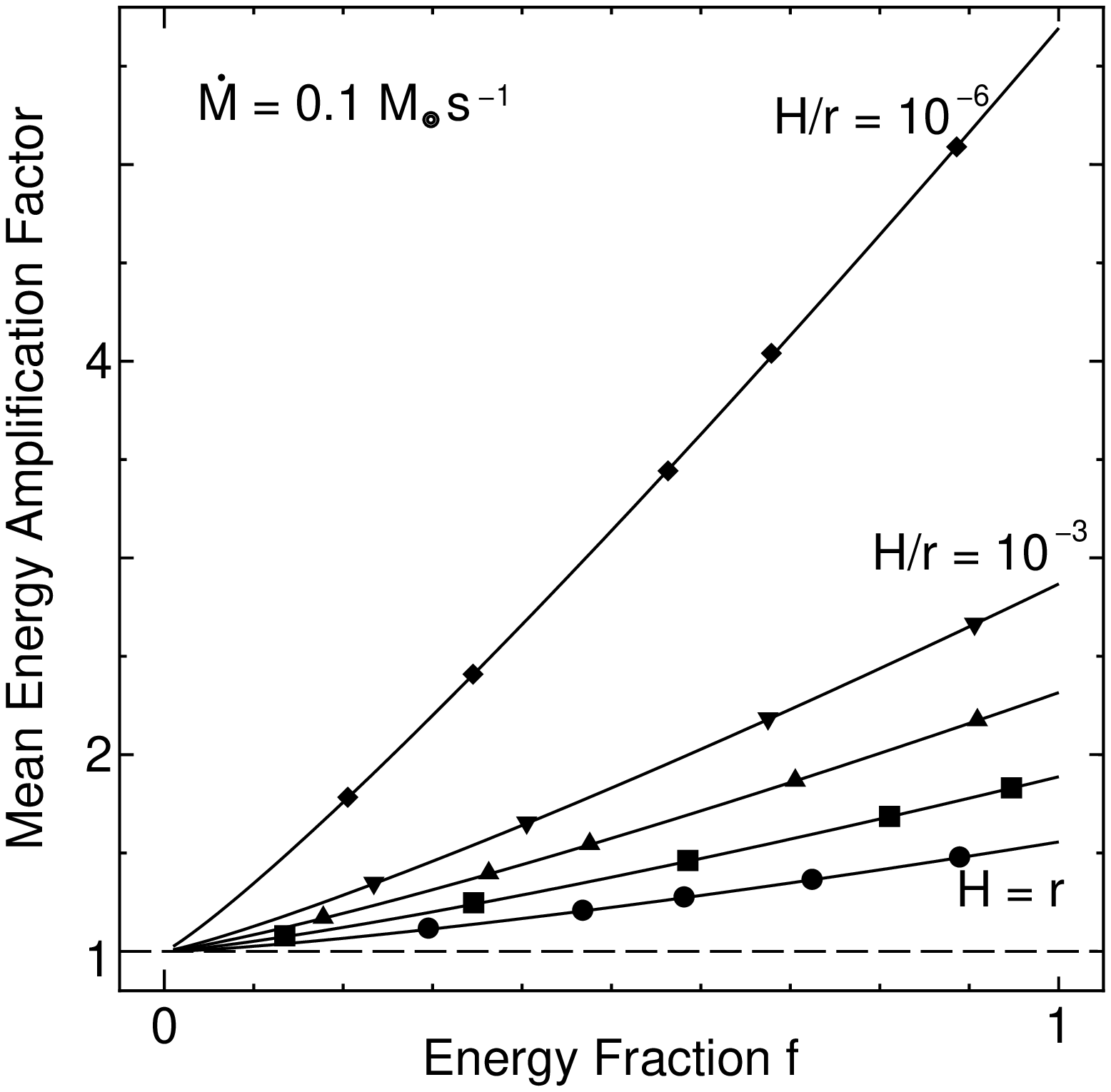}{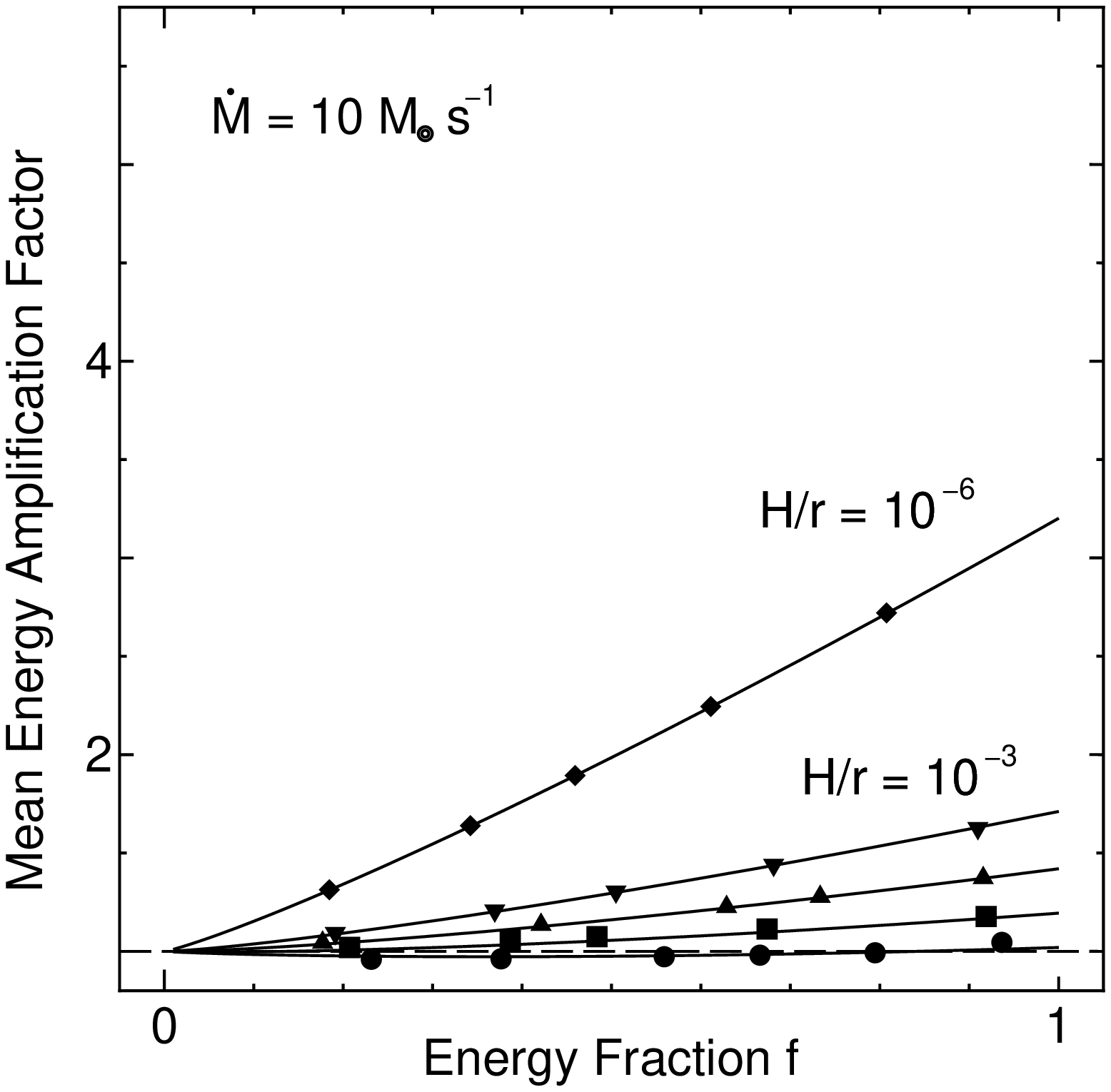}
\caption{Same as Fig.\ref{fig:spectra} ($right$) but for the case with $\dot{M} = 0.1 M_{\odot} \mathrm{s}^{-1}$ ($left$) 
and $\dot{M} = 10 M_{\odot} \mathrm{s}^{-1}$ ($right$).
In the case with $\dot{M} = 0.1 M_{\odot} \mathrm{s}^{-1}$ the neutrino mean energy is normalized by 9.7 MeV, 
and with $\dot{M} = 10 M_{\odot} \mathrm{s}^{-1}$ normalized energy is 31 MeV.
\label{fig:mdot-depend}
}
\end{figure}

\clearpage

\end{document}